\documentclass[12pt]{iopart}

\usepackage{graphicx}
\usepackage{amssymb}

\def\mb{\mu_B}   \def\tfz{T_{\rm fz}}   \def\af{\alpha} \def\bs{\bf}
\def\om{\omega}  \def\ve{\varepsilon}  
\def\tcr{T_{\rm cr}}   \def\mcr{{\mb}_{\rm cr}}  \def\pt{p_{\rm t}}
\def\ptflow{\bs p_{\rm t}^{\rm flow}}  \def\ptherm{\bs p_{\rm t}^{\rm therm}}
\def\ptvib{\bs p_{\rm t}^{\rm vib}} \def\la{\langle}   \def\d{\rm d}
\def\fc{\frac}   \def\sNN{\sqrt{s_{\rm NN}}}  \def\ra{\rangle}
   \def\bcs{\begin{cases}} \def\ecs{\end{cases}}

\begin{document}

\title[QCD phase diagram and locating
critical point using RHIC low energy scan data]{The second order
phase transition in QCD phase diagram and a new approach for
locating critical point using RHIC low energy scan data}


\author{XU Mingmei$^1$, YU Meiling$^2$,
LIU Lianshou$^1$}

\address{$^1$ Institute of Particle Physics, Huazhong Normal
University, Wuhan 430079, China\\$^2$ Department of Physics, Wuhan
University, Wuhan 430072, China} \ead{liuls@iopp.ccnu.edu.cn}

\begin{abstract}
It is shown that the RHIC energy-scan experiments can serve as an
effective tool for studying the system evolution along the first
order phase transition line passing the critical point, which is a
second order phase transition process. During this process the
system structure changes while passing the critical point, and
correspondingly, the transverse momentum of the final state
particles gets an extra component. This phenomenon can provide
useful information about the system structure in different phases
and can serve as an effective  signal for locating the critical
point.
\end{abstract}


\maketitle

Experimental evidence for the long-expected new state of matter ---
quark-gluon plasma QGP\;\cite{QGP} has been observed at the
relativistic heavy ion collider RHIC in Brookheaven National Lab
(BNL)\;\cite{evidence}. It opens a new era for the study of the
phase diagram of quantum chromodynamics QCD\;\cite{phasediagram}.

The first question is: of which order is the phase transition
between hadron gas and QGP. This has been studied
carefully\;\cite{order}. It turns out that for the realistic case of
3 massive quarks, the transition at zero chemical potential is an
analytic crossover\;\cite{order}\cite{crossover}. On the other hand,
at zero temperature the transition turns out to be of the first
order\;\cite{firstorder}. These indicate that the first order phase
transition line ends at a certain point, referred to as critical
point\;\cite{phasediagram} (temperature $\tcr$ and baryon chemical
potential $\mcr$). At even lower $\mu_B$ and higher $T$ the
transition is a smooth crossover. The first order phase transition
line and the crossover-band\;\cite{band} constitute the phase
boundary between hadron gas and QGP.

As is well known, when a system evolves along the phase boundary,
moving from the first order phase transition line to the crossover
region, it experiences a second order phase transition. During this
transition the system structure undergoes a qualitative change while
passing the critical point. The aim of the present article is to
discuss how to experimentally study such a second order phase
transition and, in particular, how to examine the change in
system-structure during this transition and how to locate the
critical point.

At BNL the RHIC experiments have started a ``low energy scan''
project\;\cite{scan} to perform the collision of two nuclei A+A,
e.g. Au+Au, at various energies lower than the top RHIC energy
$\sNN=200$ GeV. The primary goal of this project is to locate the
critical point.

Many variables have been proposed to identify the critical
point\;\cite{signals}. The basic idea of these methods is to study
the A+A collisions at different energies separately and to see
whether any one of them possess the peculiar property of critical
point, e.g. abnormally large fluctuations of some variable(s), such
as transverse momentum $\pt$, K/$\pi$ ratio, etc..

In the present article we will show that the same energy-scan data
can be used in a different, more natural, way, i.e. instead of
considering the A+A collisions at different energies separately,
take the energy-scan data as a whole and use it to study the second
order phase transition process when the system evolves along the
first order phase transition line passing through the critical point
and entering the crossover region.

\begin{figure}
\includegraphics[width=3.4in]{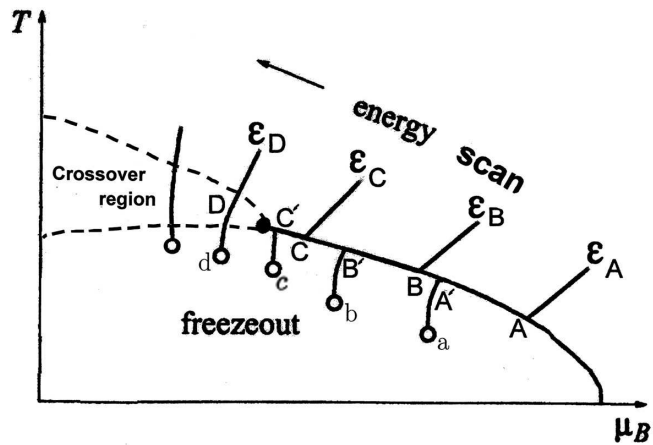}
\caption{\label{Fig. 2}  A schematic plot for the evolution of the
systems produced in the energy scan of heavy ion collisions.}
\end{figure}

In Fig.\;1 is shown a schematic plot for the evolution of the
systems produced in the energy scan of heavy ion collisions A+A at
various energies\;\cite{sketch}. After thermalization the initial
energy densities of the produced systems are $\ve_A$, $\ve_B$,
$\ve_C$, \dots. As the systems expand, temperature decreases and the
systems arrive at the phase boundary $A$, $B$, $C$ in Fig.\;1. At
this time the systems are purely partonic and a first order phase
transition starts. Since the systems are adiabatic, temperature
increases due to the release of latent heat and the systems move up
along the phase transition line~\footnote{In case of isothermal
phase transition the temperatures and chemical potentials of the two
phases will always keep equal to the critical values while the two
phases co-exist during the first order transition. The whole
transition process occurs at a single point on the $T$-$\mu$ plane.
In that case the latent heat released will be absorbed by an
external heat-bath.}. When the whole systems are converted to
hadronic phase at the points $A'$, $B'$, $C'$, they depart the phase
transition line, and freeze out to final state particles at the
points $a$, $b$, $c$. On the other hand, if the initial energy
density is as high as $\ve_D$, the system will pass through the
crossover region $D$, and freezes out at $d$.

The systems at the points $A$, $A'$, $B$, $B'$, $C$, $C'$, \dots in
Fig.\;1 are produced in A+A collisions at different $\sNN$. They are
different systems. However, since they are assumed to have arrived
thermal equilibrium, there is no difference whether they are
produced in A+A collisions at different $\sNN$ or they are coming
from the evolution of a single system through exchanging heat and
particle with an imaginary external heat- and particle-bath. In
particular, we can take the viewpoint that when the system produced
at energy $(\sNN)_A$ evolves and arrives at the point $A'$, it does
not depart the phase transition line, but continues to move along
this line through exchanging heat and particle with an ``external
bath'', and eventually arrives at $B$, $B'$, $C$, $C'$, $D$, and in
this way accomplishes a second order phase transition process.

\begin{figure}
\includegraphics[width=3.4in]{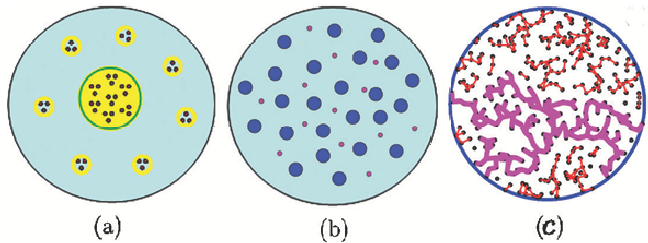}
\caption{\label{Fig. 2} (Color online) (a) The coexistence of QGP
droplet (the bigger circle at the center) and hadrons (small circles
around the bigger one) with a clear phase boundary in between. (b)
The mixture of hadrons (bigger circles) and partons (small circles)
in a gaseous state. (c) The mixture of finite-size clusters and
infinite cluster (see text).}
\end{figure}

In general, during a second order phase transition process the
system structure should undergo a qualitative change while passing
the critical point. Then what is the qualitative structure-change
when the system undergoes the above-mentioned second order phase
transition process in QCD phase diagram? In other words, how is the
system different in structure at the left and right sides of the
critical point, i.e. at crossover and first order phase transition
regions respectively? This problem has been studied in detail in
Ref.~\cite{XYL-PRL}. Let us give here a brief review.

Usually the transition from hadron phase to QGP is assumed to
proceed through the fusion of hadrons to a large region of
free-moving quarks and gluons, and the latter is conventionally
referred to as QGP-droplet, cf. Fig.\;2(a). We refer to this kind of
hadron aggregation as {\it gas-like aggregation}. It is the natural
scheme for a first order phase transition, where a coexistence of
hadronic and partonic phases with a phase boundary in between is
expected in the intermediate stage. However, when this mechanism is
applied to the analytic-crossover process~\cite{qmd, ampt}, there
will be in the intermediate stage a mixture of colored partons and
color-singlet hadrons in a gaseous state, cf. Fig.\;2(b), which
contradicts color-confinement and is, therefore, unacceptable. To
solve this problem a new form of hadron aggregation
--- {\it molecule-like aggregation} (MAM) is proposed~\cite{XYL-PRL},
where the quarks in adjacent hadrons are allowed to penetrate
through the potential barrier between the hadrons and bond the
hadrons to cluster. The hadrons in a cluster then become colored
objects, which will be referred to as {\it cells}, and only the
cluster as a whole is color-singlet. The 2-cell, 3-cell, \dots
clusters are in fact a special kind of multi-quark hadrons, which is
unallowed at zero temperature but appear at high temperature.

As temperature increases, the average size of cluster increases, and
when an infinite cluster, which in a finite system is a cluster
extending from one boundary to the other, is formed, we say that a
new phase --- QGP appears. The crossover process is then simply the
formation and growth up of color-singlet clusters and no
contradiction with color confinement any more, cf. Fig.\;2(c).

The assumption on {\it two kinds of hadron aggregation} or MAM, is
successful in providing a crossover scheme consistent with color
confinement. To have an experimental check for this assumption is
important. The RHIC energy-scan experiments can serve for this
purpose.

According to MAM when the system moves along the first order phase
transition line, passing through the critical point and entering the
crossover region, i.e. moves along $AA'BB'CC'D$ in Fig.\;1, its
structure undergoes the change from gas-like aggregation to
molecule-like aggregation, i.e. the basic element of the system
changes from single particle --- single parton at the points $A$,
$B$, $C$; and single hadron at the points $A'$, $B'$, $C'$ and $a$,
$b$, $c$ --- to cluster. As a result the cells in clusters obtain a
new vibration degree of freedom in comparison with the single
particles in gas-like aggregation case,

\begin{figure}
\includegraphics[width=1.8in]{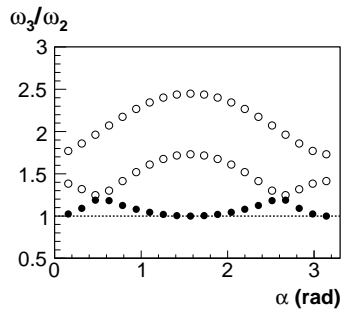}
\caption{\label{Fig. 2} The frequencies for different modes of
vibration  of a 3-cell cluster, where $\af$ is the half-angle
between the 2 bonds. The dashed line is the frequency $\om_2$ of a
2-cell cluster. The lowest frequencies for different $\af$'s are
shown as solid symbols.}
\end{figure}

The basic experimental observable in heavy ion collisions is the
momentum distribution  of final state particles. In particular, the
first order moment of transverse momentum distribution, i..e. the
average transverse momentum, versus the c.m. energy $\sNN$ is
usually used to characterize the excitation of the system. This is a
slowly increasing smooth curve, refereed to as {\it base line of
excitation} in the following discussion, which is coming from two
origins: the random thermal motion and ordered collective flow. Note
that even at the left of critical point, i.e. in the crossover
region, the vibration degree of freedom does not contribute to the
excitation base line, because the thermal excitation of vibration is
frozen.

When the system departs the crossover region with the decreasing of
temperature, the infinite cluster dissociates and the large clusters
break up to smaller ones. At the freeze-out point only small
clusters survive. Upon freeze out the small clusters break up to
single cells --- hadrons, turning the zero point vibration energy
inside clusters to the excitation energy of hadrons. Thus when the
system evolves along the first order phase transition line, passing
through the critical point and entering the crossover region, the
excitation gets an extra component, which is transformed from the
zero point vibration energy inside clusters, i.e. we have
\begin{equation} \bs p_{\rm t}= \cases { \ptflow+ \ptherm  & \mbox{\rm
when\ } $\mb > {\mb}_{\rm cr}$ , \\ \ptflow+\ptherm+\ptvib  &
\mbox{\rm when\ } $\mb < {\mb}_{\rm cr}$ , } \end{equation} where
$\ptflow$ and $\bs p_{\rm t}^{\rm therm}$ are the transverse momenta
of the frozen-out hadrons coming from the transverse expansion of
the system ({\it radial flow}) and from the thermal motion,
respectively, while $\ptvib$ denotes the excitation components
transformed from the zero point vibration inside clusters.

Let $\om_2$ be the vibration frequency of a 2-cell cluster. A 3-cell
cluster with equal length and strength of bonds and equal mass of
cells has 3 vibration modes. The corresponding frequencies are shown
in Fig.\;3\;\cite{Landau}. For energy reason the realized mode is
the lowest frequency one, shown in Fig.\;3 as solid symbols. These
frequencies are within 1 -- 1.2 $\om_2$. We assume that at the
freeze-out point all the cells posses a vibration frequency
$\bar\om$ which takes value not far from $\om_2$. The momentum
distribution corresponding to the zero-point vibration of this
frequency is
\begin{equation} \fc{\d{\rm P} (\bs p)}{\d\bs p}=
\fc{1}{(\pi m \bar\om)^{3/2}}\;\e^{-\fc{\bs
p^2}{m\bar\om}},\end{equation} where $m$ is the mass of the cell.
The vibration component of average transverse momentum $\la
\ptvib\ra$ can be calculated from this distribution.

There are at present quite a number of published $\pt$ data for
heavy ion collision at different energies. However, the colliding
nuclei and the quality of data vary widely, and it is hard to
compare the data from different experiments due to the different
systematic errors. To check the above prediction the high quality
energy-scan data with high statistics are expected.

\begin{figure}
\includegraphics[width=5in]{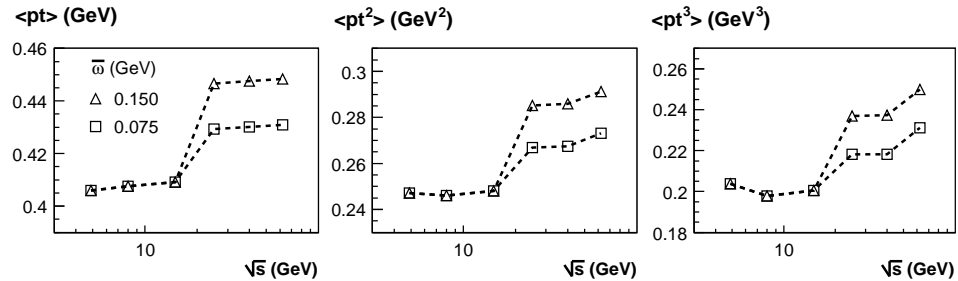}
\caption{\label{Fig. 2} The  first, second, and third order moments
of $\pt$ versus $\sNN$ for 4 values of $\bar\om$.}
\end{figure}

In order to get an idea about how large is the effect of $\ptvib$ we
simulate the ``radial flow plus thermal motion'' base line of
excitation through interpolating the experimentally fitted
parameters $\beta$ and $\tfz$\;\cite{tfz} of Au-Au collisions at
$\sNN=$ 4.88 and 62.3 GeV. Assuming that the critical point is
located in this energy region, e.g. at $\sNN=20$ GeV, the resulting
first, second and third order moments of transverse momentum are
plotted in Fig's.\;4 for 4 different values of $\bar\om$. In the
calculation the mean mass of pion, kaon and proton, averaged
according to the ratio of their yields, are used. The ordinates in
Fig's.\;4 are, therefore, the transverse-momentum moments of the
inclusive charged hadron.

It can be seen from Fig's.\;4 that a jump in $\pt$ moments at the
critical point on top of the base line is clearly visible in all
cases. For $\bar \om=75$ and 150 MeV the relative rise for the
first, second and third order moments are 4.9, 7.6, 8.8 \% and 9.2,
14.9, 18.2 \%, respectively. This effect should be observable in a
high quality data.

It can also be seen from Fig's.\;4 that the amount of moment-rising
is sensitive to the value of $\bar\om$. Therefore, if a moment
rising is observed in the energy-scan experiments, then through
fitting the data we can get the value of $\bar\om$, and, wherefrom,
obtain the information about the bond strength in the clusters,
which is a piece of useful issue in the study of QCD phase
structure.

Furthermore, the place of the $\pt$-moment-rising can be used for
locating the critical point in experiments. This method of locating
critical point has two advantages in comparison with the commonly
suggested signal of critical point\;\cite{signals} basing on the
large fluctuations at this point.

1) From thermodynamic consideration it can be shown that some
variable(s), e.g. energy density, will have large fluctuations at
the critical point. When this is used to locate the critical point,
one has to study the event-by-event fluctuations of some
variable(s), e.g. the e-by-e fluctuations of transverse momentum
$\pt$, K/$\pi$ ratio, etc.. However, since the number of particle in
a single event is finite, these {\it dynamical} fluctuations are
inevitably  contaminated by the {\it statistical} ones. Various
attempts have been made to eliminate the {\it statistical
fluctuations}\;\cite{eliminateSF}, but none of them are decisive. It
is unclear whether such a kind of fluctuation-signal for critical
point could survive after eliminating the statistical fluctuations.
On the contrary, the $\pt$ moments used in the present approach are
averaged over the whole event sample instead of event-by-event, and,
therefore, are free from the contamination of statistical
fluctuations.

2) It is obvious that in the first round the energy scan will be
performed with large steps between different colliding energies. If
it is not by occasion that some energy used in the first round just
locates at the vicinity of the fluctuation peak, we will see nothing
in the first round while measuring the fluctuations, and the
subsequent scan has to be carried out in finer steps over the whole
energy range. On the contrary, if the $\pt$ moments have a sudden
rise when passing through the critical point, then already in the
first round of energy scan we will be able to observe a rise of
these moments, provided the statistics is high and the error bars
are small. Most probably the critical point is located in the region
of the moment-rising. Then the further scan could be concentrated in
this region and the critical point, signaled by the abrupt jump of
the $\pt$ moments, can be catched in this way.

In summary, it is argued that the RHIC energy-scan data taken as a
whole can be used to study the system evolution along the first
order phase transition line, passing the critical point and entering
the crossover region. This is a second order phase transition
process. During this process the system structure changes from
gas-like to molecule-like, i.e. the basic element of the system
changes from single hadron to cluster while passing the critical
point. Correspondingly, the transverse momentum of the final state
particles get an extra component coming from the zero-point
vibration of clusters, which can be observed in experiments as an
abrupt rise of $\pt$-moments. This phenomenon can provide us useful
information about the system structure in different phases and can
serve as a signal for locating the critical point.

The proposed signal for critical point is encouraging but is not
unique. In the experimental study for locating critical point, a
combination of the signal proposed in the present article with other
signals\;\cite{signals} will provide a more decisive conclusion.

{\bf Acknowledgement} \ This work is supported by NSFC under
projects No.10775056, 10835005 and 10847131. The authors thank F.
Wang, N. Xu, Z.-B. Xu and X.-N. Wang for helpful discussions and
comments.

\end{document}